\documentclass[11pt,graphicx,amsmath]{article}
\usepackage{amsmath}
\usepackage{graphicx}
\usepackage{bm}
\usepackage[dvips]{color}
\usepackage{amssymb}
\usepackage{amsfonts}
\usepackage{comment}
\usepackage{cite}
\usepackage{todonotes}
\usepackage{caption}
\usepackage{subcaption}

\def\be{\begin{equation}}
\def\ee{\end{equation}}
\def\nn{\nonumber}

\def\ba{\begin{eqnarray}}
\def\ea{\end{eqnarray}}
\def\bl#1\el{\begin{align}#1\end{align}}

\def\bl#1\el{\begin{align}#1\end{align}}

\title{The nonlinear field equation of the three-point correlation function of galaxies}

\author{\small        Shu-Guang  Wu\thanks{wusg@mail.ustc.edu.cn}, \,
                         Yang  Zhang\thanks{yzh@ustc.edu.cn},  \\
 \small  Department of  Astronomy,  Key Laboratory
               for Researches in Galaxies and Cosmology, \\
 \small    School of Astronomy and Space Sciences, \\
 \small  University of Science and Technology of China,  Hefei, Anhui, 230026,  China \\
 }

 \date{}

\date{}

\def\be{\begin{equation}}
	\def\ee{\end{equation}}
\def\ba{\begin{eqnarray}}
	\def\ea{\end{eqnarray}}
\def\nn{\nonumber}
\def\la{\langle}
\def\ra{\rangle}

\allowdisplaybreaks

\sf

\begin{document}
	
\maketitle
	
\begin{abstract}

Based on the field theory of density fluctuation under Newtonian gravity,
we obtain analytically the nonlinear equation of
3-pt correlation function   $\zeta$ of galaxies
in a homogeneous,  isotropic,  static universe.
The density fluctuations have been kept up to second order.
By the Fry-Peebles ansatz and the Groth-Peebles ansatz,
the equation of $\zeta$  becomes closed
and differs from the Gaussian approximate equation.
Using the boundary condition inferred from the data of  SDSS,
we obtain the solution $\zeta(r, u, \theta)$ at fixed $u=2$,
which exhibits a shallow $U$-shape along the angle $\theta$
and, nevertheless, decreases monotonously along  the radial $r$.
We show its difference with the Gaussian solution.
As a direct criterion of non-Gaussianity,
the reduced $Q(r, u, \theta)$ deviates from the Gaussianity plane $Q=1$,
exhibits a deeper $U$-shape along  $\theta$ and varies  weakly along $r$,
agreeing  with the observed data.

\end{abstract}

{\bf keywords:}
  cosmology: large-scale structure of Universe
  -cosmology: theory
-gravitation

-hydrodynamics

\

\section{Introduction}
\label{Intro}

The n-point correlation functions  are important tool to
study the statistical properties of matter distribution on the large scale of the universe
and can provide fundamental tests of the standard cosmological model
\cite{Peebles1980,Bernardeau2002}.
The statistic of noninterating particles, like CMB,
can be well described as statistically Gaussian random field,
the 2-point correlation function (2PCF) will be sufficient
to characterize its correlation.
When long-range Newtonian gravity is taken into account,
the concept of a Gaussian random field has been subtle in literature so far.
Therefore, a criterion of non-Gaussianity is required
to be defined clearly.
The equation of 2PCF $G^{(2)}(r)$ of density fluctuation  to lowest order
under Newtonian gravity is a Helmholtz equation with a delta source,
and the exact solution has been given
and called the solution in the Gaussian approximation in Ref.\cite{Zhang2007}.
This is because the equation $G^{(2)}(r)$ shares a structure similar to
 the Gaussian approximate equation \cite{Goldenfeld1992}
that has been commonly used in condensed matter physics.
Parallelly,  the equation of  3-pt correlation function (3PCF)
  of density fluctuation to lowest order (the Gaussian approximation)
is also a linear equation and the exact solution  \cite{ZhangChenWu2019}
has been found as the following
\be \label{gauss3pt}
G^{(3)}(r_{12},r_{23},r_{31})
= Q  \big[G^{(2)}(r_{12}) G^{(2)}(r_{23})
 + G^{(2)}(r_{23}) G^{(2)}(r_{31})
 +G^{(2)}(r_{31}) G^{(2)}(r_{12}) \big]
\ee
where $Q=1$,  and  $r_{12}=|{\bf r}_1 - {\bf r}_2|$,  etc.
Thus, $Q=1$ holds in the Gaussian approximation,
and any deviation of $Q$ from 1 will be an indication of
non-Gaussianity of the  density fluctuation.
Interestingly,
the solution \eqref{gauss3pt} in  the Gaussian approximation
is exactly the content of the Groth-Peebles ansatz with $Q=1$
\cite{GrothPeebles1975,GrothPeebles1977}.
When density fluctuations up to second order are included,
the equations of $G^{(2)}(r)$ becomes nonlinear
\cite{zhang2009nonlinear,ZhangChen2015,ZhangChenWu2019},
and its solution describes the distribution of galaxies
better than the Gaussian approximation at small scales.
But  $G^{(3)}$ has not been analytically studied
 up to second order of density fluctuation.
Statistically, $G^{(3)}(\mathbf{r, r', r''})$
 describes the excess probability over random of
finding three galaxies located at the three vertices (${\bf r}$, ${\bf r}'$, ${\bf r}''$)
of a given triangle.
In  observations and numerical studies,
as an extension of the Groth-Peebles ansatz \eqref{gauss3pt},
 the reduced 3PCF  is often  introduced
\be \label{grothpeeblesansatz}
Q(\mathbf{r, r', r''})
 \equiv \frac{G^{(3)}(\mathbf{r, r', r''})}
 {G^{(2)}(\mathbf{r, r'})  G^{(2)}(\mathbf{r', r''})
+G^{(2)}(\mathbf{r', r''}) G^{(2)}(\mathbf{r'', r})
+G^{(2)}(\mathbf{r'', r}) G^{(2)}(\mathbf{r, r'})},
\ee
As a direct criterion,
$Q(\mathbf{r, r', r''})$ indicates the non-Gaussianity when it deviates from $1$.
Galaxy surveys show that $Q \ne 1$,
and confirm the non-Gaussianity of the distribution of galaxies.
Moreover,  $Q$ depends on the scale and shape of the triangle
\cite{Jing1998,Nichol2006,Gaztanaga2009, marin2011, McBride2011a,
McBride2011b, Guo2016, Slepian2017,Jing2004, Wang2004, Gaztanaga2005},
a feature also occurring in simulations
   \cite{FryMellot1993,BarrigaGaztanaga2002,GaztanagaScoccimarro2005}
and in the study by perturbation theory \cite{Fry1994,Bernardeau2002}.

In this paper, as a continuation of serial study
\cite{Zhang2007,zhang2009nonlinear,ZhangChen2015,ZhangChenWu2019},
we shall derive analytically
the nonlinear  field equation of $G^{(3)}$  up to second order of density fluctuation
beyond Gaussian approximation,
give the solution $G^{(3)}$.
As have been shown \cite{ZhangLi2021},
the evolution effect of correlation function of galaxies
is not drastic within a low redshift range  ($z= 0.5 \sim 0.0$),
so for simplicity we study the nonevolution case
and compare with observations ($z= 0.16 \sim 0.47$) \cite{marin2011}
in this preliminary work,
and the evolution case will be given in future.

\section{Nonlinear Field Equation of 3-point Correlation Function}
\label{sec:deri3PCF}

Within the framework of Newtonian gravity,
the distribution of galaxies and clusters in a static Universe
can be described by the density field  $\psi$
with the equation \cite{Zhang2007,zhang2009nonlinear,ZhangChen2015,ZhangChenWu2019}
\be \label{psifieldequ}
\nabla^2 \psi-\frac{(\nabla \psi)^2}{\psi}+k_J^2 \psi^2+J \psi^2 =0 ,
\ee
where $\psi(\mathbf{r}) \equiv  \rho(\mathbf{r}) / \rho_0$
is the rescaled mass density with $\rho_0$ being the mean mass density,
and  $k_J \equiv (4\pi G \rho_0/c_s^2)^{1/2}$ is the Jeans wavenumber,
$c_s $ is the sound speed,
and the source $J$ is used to handle the functional derivatives with ease.
The generating functional for the correlation functions of $\psi$ is given by
\be  \label{Zdef}
Z [J]= \int D \phi \exp[-\alpha \int {\rm d}^3 {\bf r} \mathcal{H}(\psi, J)],
\ee
where  $\alpha  \equiv   c_s^2/ 4\pi G m$ and the effective Hamiltonian is
\be  \label{effL}
\mathcal{H}(\psi, J)
 = \frac{1}{2} \bigg(\frac{\nabla \psi}{\psi}\bigg)^2 - k_J^2 \psi - J \psi .
\ee
The connected $n$-point correlation function of $\delta \psi$  is
\bl \label{npcf}
G^{(n)}({\bf r}_1,\cdots,{\bf r}_n)
& =\la \delta \psi({\bf r}_1) \cdots \delta \psi({\bf r}_n)\ra
\nn
\\
& =\frac{1}{\alpha^n}\frac{\delta^n \log Z[J]}{\delta J({\bf r}_1)
\cdots\delta J({\bf r}_n)}\vert_{J=0}
=\frac{1}{\alpha^{n-1}}\frac{\delta^{n-1} \la \psi({\bf r}_1) \ra }
{\delta J({\bf r}_2)\cdots\delta J({\bf r}_n)}\vert_{J=0},
\el
where $\delta \psi({\bf r}) = \psi({\bf r}) - \la \psi({\bf r}) \ra$
is the fluctuation field
around the expectation value $\la \psi({\bf r}) \ra $.
(See Refs.\cite{Goldenfeld1992,Zhang2007,zhang2009nonlinear,ZhangChen2015,ZhangChenWu2019}.)
To derive the field equation of
the 3-point correlation function $G^{(3)}({\bf r}, {\bf r}', {\bf r}'')$,
we take the ensemble average of Eq.(\ref{psifieldequ}) in the presence of $J$,
and take the functional derivative of this equation
twice with respect to the source $J$, and set $J=0$.
In calculation,
the second term in Eq.(\ref{psifieldequ}) is approximated by
\be \label{secondterm}
\la \frac{(\nabla \psi)^2}{\psi} \ra
\approx \frac{(\nabla \la \psi \ra)^2}{\la \psi \ra}
+ \frac{\la (\nabla \delta \psi)^2 \ra}{\la \psi \ra}
-\frac{\nabla \la \psi \ra}{\la \psi \ra^2}  \cdot \la \nabla (\delta \psi)^2 \ra
+\frac{(\nabla \la \psi \ra)^2}{\la \psi \ra^3} \la (\delta \psi)^2 \ra,
\ee
where the  second order fluctuation    $(\delta \psi)^2$ is kept
and higher order  terms have been neglected.
By lengthy  and straightforward calculations,
using the definition (\ref{npcf}),
we obtain  the field equation of $G^{(3)}(\mathbf{r, r', r''})$
 up to the second order of density fluctuation
as the following
\ba \label{3PCF}
&& \nabla^2 G^{(3)}(\mathbf{r, r', r''})
+\frac{2}{\psi_0^2}  \nabla G^{(2)}(0) \cdot \nabla  G^{(3)}(\mathbf{r, r', r''})
+\bigg(2 k_J^2 \psi_0+\frac{1}{2 \psi_0^2} \nabla^2  G^{(2)}(0) \bigg) G^{(3)}(\mathbf{r, r', r''})
\nonumber \\
& &+\frac{1}{2 \psi_0^2}  G^{(2)}(\mathbf{r, r'}) \nabla^2 G^{(3)}(\mathbf{r, r, r''})
+\frac{1}{2 \psi_0^2} G^{(2)}(\mathbf{r, r''}) \nabla ^2 G^{(3)}(\mathbf{r, r, r'})  \nonumber \\
&&+\frac{2}{\psi_0^2}\nabla  G^{(2)}(\mathbf{r, r'}) \cdot  \nabla G^{(3)}(\mathbf{r, r, r''})
+\frac{2}{\psi_0^2} \nabla G^{(3)}(\mathbf{r, r, r'}) \cdot \nabla G^{(2)}(\mathbf{r, r''}) \nonumber \\
& &-\frac{1}{2 \psi_0} \nabla ^2  G^{(4)}(\mathbf{r, r, r', r''})
- k_J^2 G^{(4)}(\mathbf{r, r, r', r''}) \nonumber \\
&&- \frac{2}{\psi_0} \bigg( \frac{2}{\psi_0^2} G^{(2)}(0)
+1 \bigg) \nabla G^{(2)}(\mathbf{r, r'}) \cdot \nabla G^{(2)}(\mathbf{r, r''}) \nonumber \\
& &
+\bigg( 2 k_J^2 -\frac{1 }{\psi_0^3} \nabla^2  G^{(2)}(0) \bigg) G^{(2)}(\mathbf{r, r'}) G^{(2)}(\mathbf{r, r''}) \nonumber \\
& &
-\frac{4}{\psi_0^3} G^{(2)}(\mathbf{r, r''}) \nabla G^{(2)}(0) \cdot \nabla  G^{(2)}(\mathbf{r, r'})
-\frac{4}{\psi_0^3} G^{(2)}(\mathbf{r, r'}) \nabla G^{(2)}(0) \cdot \nabla G^{(2)}(\mathbf{r, r''}) \nonumber \\
&=&\frac{1}{\alpha} \delta^{(3)}(\mathbf{r-r'}) G^{(3)}(\mathbf{r, r, r''})
+\frac{1}{\alpha} G^{(3)}(\mathbf{r, r, r'}) \delta^{(3)}(\mathbf{r-r''})\nonumber \\
&&-\frac{2 \psi_0 }{\alpha }  \delta^{(3)}(\mathbf{r-r'})   G^{(2)}(\mathbf{r, r''})
-\frac{2 \psi_0 }{ \alpha } \delta^{(3)}(\mathbf{r-r''}) G^{(2)}(\mathbf{r, r'}),
\ea
where $ G^{(2)}(0) \equiv G^{(2)}(\mathbf{r, r})$
and $\psi_0 \equiv  \la \psi({\bf r}) \ra_{J=0}=1$.
When the higher order terms, such as $ G^{(2)} G^{(3)}$ and $ G^{(4)}$,   are dropped,
 Eq.(\ref{3PCF}) reduces to that of
the Gaussian approximation.
 (See Eq.(28) in Ref. \cite{ZhangChenWu2019}).

Yet, Eq.(\ref{3PCF}) is  not closed for $G^{(3)}$,
as it hierarchically contains the higher order 4-point correlation function
$G^{(4)}$ terms.
To deal with it, we adopt Fry-Peebles ansatz  \cite{FryPeebles} as the following
\ba \label{frypeeblesansatz}
G^{(4)}(\mathbf{r_1, r_2, r_3, r_4})
&=&R_a[ G^{(2)}(\mathbf{r_1, r_2}) G^{(2)}(\mathbf{r_2, r_3}) G^{(2)}(\mathbf{r_3, r_4}) +\mathrm{sym. (12 \, \, terms)} ] \nonumber \\
& &+R_b[ G^{(2)}(\mathbf{r_1, r_2}) G^{(2)}(\mathbf{r_1, r_3}) G^{(2)}(\mathbf{r_1, r_4})+\mathrm{sym. (4 \, \, terms)} ],
\ea
where $R_a$ and $R_b$ are constants,
$R_a$ and $R_b$ around $1 \sim 10$  roughly
\cite{Fry1983,Fry1984,Szapudi1992, Meiksin1992,Peebles1993}.
By the ansatz \eqref{frypeeblesansatz},
the  $G^{(4)}$ term  in  Eq.(\ref{3PCF}) is written as
\ba \label{g4rr}
&&G^{(4)}(\mathbf{r, r, r', r''})
=2 R_a \big( G^{(2)}(\mathbf{r, r'})
+ G^{(2)}(\mathbf{r, r''}) \big) \big( G^{(2)}(0) G^{(2)}(\mathbf{r', r''})
+ G^{(2)}(\mathbf{r, r'}) G^{(2)}(\mathbf{r, r''}) \big)
\nonumber \\
& &+2 (R_a + R_b) G^{(2)}(0) G^{(2)}(\mathbf{r, r'}) G^{(2)}(\mathbf{r, r''})
+2 R_a G^{(2)}(\mathbf{r, r'}) G^{(2)}(\mathbf{r, r''}) G^{(2)}(\mathbf{r', r''})  \nonumber \\
& &+R_b \big( G^{(2)}(\mathbf{r, r'})^2
+G^{(2)}(\mathbf{r, r''})^2 \big) G^{(2)}(\mathbf{r', r''}).
\ea
Eq.\eqref{3PCF} also contains the the squeezed  3PCF,
\[
G^{(3)}(\mathbf{r, r, r'}) = \frac{1}{\alpha}\frac{\delta}{\delta J({\bf r}')}
\big( \la \delta \psi({\bf r}) \delta \psi({\bf r}) \ra \big)\vert_{J=0}
\, ,
\]
which is the limit
$G^{(3)}(\mathbf{r, r, r'})=\lim\limits_{{\bf r}''
\rightarrow {\bf r}}G^{(3)}(\mathbf{r, r', r''}) $    \cite{Yuan2017}.
When ${\bf r}'' \rightarrow {\bf r}$,
the two galaxies separated by a distance $\lvert \mathbf{r''}-\mathbf{r} \rvert$
will interact  strongly via gravity,
and  $G^{(3)}(\mathbf{r, r, r'})$ will  mask or distort the signals
in observations and simulations.
Some binning schemes are often used to avoid this  difficulty
\cite{Gaztanaga2005,  McBride2011a, McBride2011b, Slepian2017}.
Ref.\cite{Yuan2017} treated the squeezed  3PCF
as a function of the pair-galaxy bias,
independent of $\lvert \mathbf{r} - \mathbf{r'} \rvert$.
However, observations indicate  that the squeezed   3PCF $Q$ depends on scale.
Here we adopt the  Groth-Peebles ansatz \cite{GrothPeebles1977}
\be \label{g3rrr'}
G^{(3)}(\mathbf{r, r, r'})
   =2 Q G^{(2)}(0) G^{(2)}(\mathbf{r, r'})+Q G^{(2)}(\mathbf{r, r'})^2,
\ee
where $Q$ is a constant
and will be treated as a new parameter in the equation of 3PCF.
Substituting (\ref{g4rr}) and (\ref{g3rrr'})  into
Eq.(\ref{3PCF}) gives the closed field equation of the 3PCF
\ba \label{3PCF_02}
&& \nabla^2 G^{(3)}(\mathbf{r, r', r''})
+{\bf a} \cdot \nabla  G^{(3)}(\mathbf{r, r', r''})
+2 g k_J^2 G^{(3)}(\mathbf{r, r', r''})
 -  \mathcal{A}(\mathbf{r, r', r''})
\nonumber \\
&=&\frac{1}{\alpha}  \Big( 2 ( Q b  -1 )+ Q G^{(2)}(\mathbf{r, r''}) \Big)
       G^{(2)}(\mathbf{r, r''}) \delta^{(3)}(\mathbf{r-r'}) \nonumber \\
&&+\frac{1}{\alpha}  \Big( 2 ( Q b -1 ) + Q G^{(2)}(\mathbf{r, r'}) \Big)
        G^{(2)}(\mathbf{r, r'}) \delta^{(3)}(\mathbf{r-r''}),
\ea
where ${\bf a} \equiv \frac{2}{\psi_0^2} \nabla G^{(2)}(0)$,
$b \equiv  \frac{1}{\psi_0^2} G^{(2)}(0)$,
$g \equiv   ( 1  + \frac{1}{4  \psi_0 k_J^2} c )$
with $c \equiv  \frac{1}{\psi_0^2} \nabla^2 G^{(2)}(0)$
are three parameters,  and
\ba \label{Adef}
&&\mathcal{A}(\mathbf{r, r', r''})  \nonumber \\
&=&   2 \big[ \big( R_a + R_b -4 Q  + 2 \big) b + 1 \big]
\nabla G^{(2)}(\mathbf{r, r'}) \cdot \nabla G^{(2)}(\mathbf{r, r''})
\nonumber \\
&& -  \big[ 2 k_J^2 -2 k_J^2 (R_a + R_b) b
-\big( R_a + R_b - 2Q + 1 \big) c
\big] G^{(2)}(\mathbf{r, r'}) G^{(2)}(\mathbf{r, r''})
\nonumber \\
&&  + \big( R_a + R_b- Q \big) b  \bigg( \nabla^2 G^{(2)}(\mathbf{r, r'})
G^{(2)}(\mathbf{r, r''})
+G^{(2)}(\mathbf{r, r'}) \nabla^2 G^{(2)}(\mathbf{r, r''}) \bigg)
\nonumber \\
&& + \big( R_a + R_b -3Q +2 \big) {\bf a} \cdot \nabla
\big( G^{(2)}(\mathbf{r, r'}) G^{(2)}(\mathbf{r, r''}) \big)
\nonumber \\
& & + \big( 2 R_a - Q \big) G^{(2)}(\mathbf{r, r'}) G^{(2)}(\mathbf{r, r''})
\big(\nabla^2 G^{(2)}(\mathbf{r, r'})
+ \nabla^2 G^{(2)}(\mathbf{r, r''}) \big)\nonumber \\
&& + \big( 4 R_a - 4 Q \big) \big(G^{(2)}(\mathbf{r, r'})
+G^{(2)}(\mathbf{r, r''}) \big) \nabla G^{(2)}(\mathbf{r, r'})
\cdot \nabla G^{(2)}(\mathbf{r, r''})  \nonumber \\
&& + \big( 2 R_a - Q \big) \bigg( G^{(2)}(\mathbf{r, r'})
\lvert \nabla G^{(2)}(\mathbf{r, r''}) \rvert^2
+ G^{(2)}(\mathbf{r, r''}) \lvert \nabla G^{(2)}(\mathbf{r, r'}) \rvert^2 \bigg) \nonumber \\
&& + R_a \bigg( G^{(2)}(\mathbf{r, r'})^2 \nabla^2 G^{(2)}(\mathbf{r, r''})
+G^{(2)}(\mathbf{r, r''})^2 \nabla^2 G^{(2)}(\mathbf{r, r'}) \bigg) \nonumber \\
& & + 2 R_a  k_J^2 \big( G^{(2)}(\mathbf{r, r'})
+ G^{(2)}(\mathbf{r, r''}) \big) G^{(2)}(\mathbf{r, r'}) G^{(2)}(\mathbf{r, r''}) \nonumber \\
& & + R_a G^{(2)}(\mathbf{r', r''}) \bigg( c G^{(2)}(\mathbf{r, r'})
+{\bf a} \cdot \nabla G^{(2)}(\mathbf{r, r'})
+b \nabla^2 G^{(2)}(\mathbf{r, r'}) \nonumber \\
&&+ c G^{(2)}(\mathbf{r, r''})
+{\bf a} \cdot \nabla G^{(2)}(\mathbf{r, r''})
+b \nabla^2 G^{(2)}(\mathbf{r, r''})  \nonumber \\
&&+\nabla^2 G^{(2)}(\mathbf{r, r'}) G^{(2)}(\mathbf{r, r''})
+2 \nabla G^{(2)}(\mathbf{r, r'}) \cdot \nabla G^{(2)}(\mathbf{r, r''})
+G^{(2)}(\mathbf{r, r'}) \nabla^2 G^{(2)}(\mathbf{r, r''}) \bigg) \nonumber \\
& & + R_b G^{(2)}(\mathbf{r', r''}) \bigg( G^{(2)}(\mathbf{r, r'}) \nabla^2 G^{(2)}(\mathbf{r, r'})
+ \vert \nabla G^{(2)}(\mathbf{r, r'}) \vert^2
+ G^{(2)}(\mathbf{r, r''}) \nabla^2 G^{(2)}(\mathbf{r, r''})
+ \vert \nabla G^{(2)}(\mathbf{r, r''}) \vert^2  \bigg) \nonumber \\
&& +  k_J^2 \bigg( 2 R_a b G^{(2)}(\mathbf{r, r'})
+2 R_a b G^{(2)}(\mathbf{r, r''})  \nonumber \\
& &+R_b G^{(2)}(\mathbf{r, r'})^2
+R_b G^{(2)}(\mathbf{r, r''})^2
+2 R_a G^{(2)}(\mathbf{r, r'}) G^{(2)}(\mathbf{r, r''}) \bigg) G^{(2)}(\mathbf{r', r''}) .
\ea
The function $\mathcal{A}$ depends on  $ G^{(2)}$.
The structure of eq.(\ref{3PCF_02}) is similar to
that in the Gaussian approximation  \cite{ZhangChenWu2019},
but contains an extra convection term
$\mathbf{a} \cdot \nabla  G^{(3)}(\mathbf{r, r', r''})$.
The 2PCF  $G^{(2)}(\mathbf{r, r'})$
has been solved up to second order of density fluctuation
\cite{zhang2009nonlinear,ZhangChen2015,ZhangChenWu2019}.
In this paper, to be consistent with observation,
we shall use the observed $G^{(2)}(\mathbf{r, r'})$ from Ref.\cite{marin2011}.
There are eight parameters $\mathbf{a}$, $b$, $c$, $g$, $Q$,
$R_a$, $R_b$ and $k_J$ in Eqs. (\ref{3PCF_02}) (\ref{Adef}),
treated as being independent,
which differ from
those in Refs.\cite{Zhang2007,zhang2009nonlinear,ZhangChen2015,ZhangChenWu2019}
in renormalization.

\section{The  solution of 3PCF equation }
\label{sec:sol3PCF}

\begin{figure}[htbp]
	\centering
	\includegraphics[width=0.5\columnwidth]{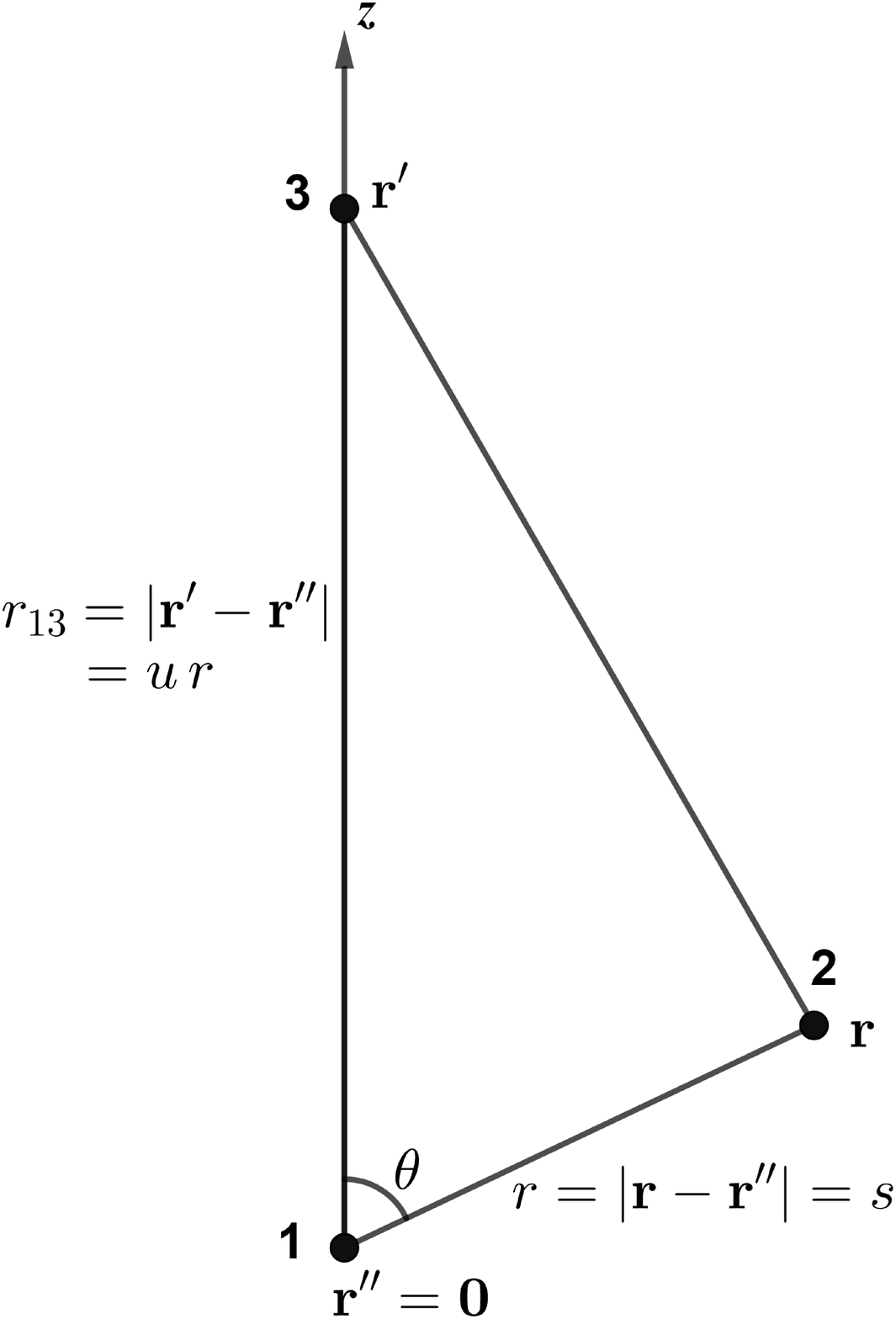}\\
	\caption{  \label{zetaconfig}
		The configuration of the triangle of $G^{(3)}(\mathbf{r, r', r''})$
		in the spherical coordinate.
Here we take the azimuth angle $\phi=0$,
		 $\mathbf{r''}=\mathbf{0}$ as the origin,
        and the vector $\mathbf{r'}-\mathbf{r''}$ along with the $z$-axis.
    	}
\end{figure}

In a homogeneous and isotropic universe,
it is assumed that $G^{(2)}(\mathbf{r, r'})= G^{(2)}(\mathbf{|r- r'|})$
and that $G^{(3)}(\mathbf{r, r', r''})$  depends only on the configuration
of a triangle with three vertexes located at $(\mathbf{r, r', r''})$.
So,   $G^{(3)}(\mathbf{r, r', r''})$ has  only three independent variables,
and is commonly  parametrized by   \cite{marin2011}
\be
G^{(3)}(\mathbf{r, r', r''}) \equiv  \zeta(s, u, \theta ) ,
\ee
where the three variables are defined as
\[
s= r_{12} \equiv r,
~~~ u=\frac{r_{1 3}}{r_{12}},
~~~~ \theta =\cos^{-1}(\hat{{\bf r}}_{12}\cdot\hat{{\bf r}}_{13}) ,
\]
which are demonstrated in Fig.\ref{zetaconfig}.
Then eq. (\ref{3PCF_02}) is written  in spherical coordinate as
\ba \label{3PCF_02sp}
&&\frac{1}{r^2}  \frac{\partial}{\partial r}
   \big(r^2 \frac{\partial}{\partial r}\zeta(r, u, \theta) \big)
+\frac{1}{r^2 \sin \theta}\frac{\partial}{\partial \theta}
      \big(\sin \theta \frac{\partial \zeta(r, u, \theta)}{\partial \theta} \big)
+a_r \frac{\partial \zeta(r, u, \theta)}{\partial r}
+2 g k_J^2 \zeta(r, u, \theta)
 - \mathcal{A}(r, u, \theta) \nonumber \\
&=& \frac{1}{\alpha}  \Big( 2 ( Q b  -1 )+ Q \xi(r) \Big) \xi(r) \delta^{(3)}(l)
+\frac{1}{\alpha}  \Big( 2 ( Q b -1 ) + Q \xi(l) \Big) \xi(l) \delta^{(3)}(r),
\ea
where  $\xi(r) \equiv G^{(2)}(|\bf r|)$,
 $a_r$  is the radial component of the vector parameter $\mathbf{a}$,
\[
l \equiv \lvert \mathbf{r}- \mathbf{r'} \rvert
  =r \sqrt{1+u^2 -2u \cos \theta}
  \equiv \beta r,
\]
and
\bl  \label{Adefsp}
 \mathcal{A}(r, u, \theta) &  =
  2 \Big[ \big( R_a + R_b -4 Q  + 2 \big) b + 1 \Big] \beta \xi'(l) \xi'(r)
   + \big( 4 R_a - 4 Q \big) \big(\xi(l) + \xi(r) \big) \beta \xi'(l) \xi'(r)
\nonumber \\
& - \Big[ 2 k_J^2 -2 k_J^2 (R_a + R_b) b
    -\big( R_a + R_b - 2Q + 1 \big) c \Big] \xi(l) \xi(r)
\nonumber \\
& +\big( R_a + R_b -3Q +2 \big) a_r \Big[ \beta \xi'(l) \xi(r)+ \xi'(r) \xi(l) \Big]
\nonumber \\
& \big( R_a + R_b- Q \big) b
\bigg[ \bigg( \big( \frac{2}{r} \beta +\frac{2 u}{\beta r}\cos \theta
          -\frac{u^2 \sin^2 \theta }{\beta^3 r} \big) \xi'(l)
+\big( \beta^2 + \frac{u^2}{\beta^2} \sin^2 \theta \big) \xi''(l) \bigg) \xi(r)
\nn \\
& +\big( \frac{2}{r} \xi'(r) + \xi''(r) \big) \xi(l) \bigg]
\nn \\
& +\big( 2 R_a - Q \big) \bigg[ \xi(l) \xi(r)
\bigg( \big( \frac{2}{r} \beta +\frac{2 u}{\beta r}\cos \theta
 -\frac{u^2 \sin^2 \theta }{\beta^3 r} \big) \xi'(l)
 +\big( \beta^2 + \frac{u^2}{\beta^2} \sin^2 \theta \big) \xi''(l)
 + \frac{2}{r} \xi'(r) + \xi''(r) \bigg)
\nonumber \\
& + \xi(l) \xi'(r)^2
 + \xi(r) \big(\beta^2 + \frac{u^2 \sin^2 \theta}{\beta^2}\big)\xi'(l)^2  \bigg] \nonumber \\
& +R_a \bigg[2 \beta \xi'(l) \xi'(r) \xi(u \, r)
+ \big( \frac{2}{r} \xi'(r) + \xi''(r)  \big) \big( \xi(l)+ \xi(u \, r) \big) \xi(l) \nonumber \\
& + \bigg( \big( \frac{2}{r} \beta +\frac{2 u}{\beta r}\cos \theta
-\frac{u^2 \sin^2 \theta }{\beta^3 r} \big) \xi'(l)
+\big( \beta^2 + \frac{u^2}{\beta^2} \sin^2 \theta \big) \xi''(l) \bigg)
\big( \xi(r)+ \xi(u \, r) \big) \xi(r) \bigg] \nonumber \\
& +2 R_a  k_J^2  \big( \xi(l) + \xi(r) + \xi(u \, r) \big) \xi(l) \xi(r)
  +R_a \big( c + 2 k_J^2 b  \big) \xi(u \, r) \big( \xi(l) + \xi(r) \big)
\nonumber \\
& +R_a a_r \xi(u \, r) \big( \beta \xi'(l) + \xi'(r) \big)
  +R_b k_J^2 \xi(u \, r) \big( \xi(l)^2+\xi(r)^2  \big)\nonumber \\
& +R_a b \xi(u \, r)  \bigg( \big( \frac{2}{r} \beta +\frac{2 u}{\beta r}\cos \theta
-\frac{u^2 \sin^2 \theta }{\beta^3 r} \big) \xi'(l)
+\big( \beta^2 + \frac{u^2}{\beta^2} \sin^2 \theta \big) \xi''(l)
+ \frac{2}{r} \xi'(r) + \xi''(r) \bigg) \nonumber \\
&
+ R_b \xi(u \, r)
\bigg[ \xi(l) \bigg( \big( \frac{2}{r} \beta +\frac{2 u}{\beta r}\cos \theta
-\frac{u^2 \sin^2 \theta }{\beta^3 r} \big) \xi'(l)
+\big( \beta^2 + \frac{u^2}{\beta^2} \sin^2 \theta \big) \xi''(l) \bigg)
\nonumber \\
&
+ \big(\beta^2 + \frac{u^2 \sin^2 \theta}{\beta^2}\big)\xi'(l)^2
+ \xi(r) \big( \frac{2}{r} \xi'(r) + \xi''(r) \big)
+ \xi'(r)^2  \bigg].
\el
Eq.\eqref{3PCF_02sp} of $\zeta(r, u, \theta)$ in spherical coordinate
will   be solved  in actual computation.
The ratio  $u=2$ is often taken in simulations
and presentations of observational data,
so that $\zeta(r, u, \theta)$ has only two variables.
We also take this in the following.

\begin{figure}[htb]
\centering
 \includegraphics[width = .95\linewidth]{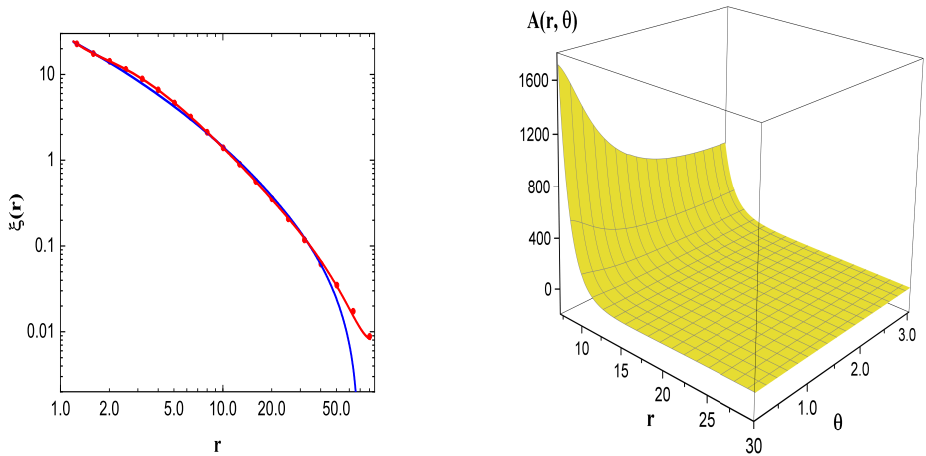}
\caption{ (a): the observed $\xi(r)$ (red with dots) from Ref.\cite{marin2011},
               the solution $\xi$ to second order (blue) from Ref.\cite{ZhangChen2015}.
          (b): $\mathcal{A}(r, u, \theta)$ in eq.\eqref{Adefsp} at fixed $u=2$
                 as function of $(r,   \theta)$.
}
\label{A}
\end{figure}

To solve the \eqref{3PCF_02sp} for $\zeta$, we need the 2PCF  $\xi(r)$.
For a coherent comparison  with   observation,
we shall use the observed $\xi(r)$ given in Figure 5 of Ref.\cite{marin2011}.
We plot Fig.\ref{A} (a) to show
the observed $\xi(r)$ (red with dots) from Ref.\cite{marin2011},
and  the nonlinear solution $\xi(r)$ (blue) from Ref.\cite{ZhangChen2015}.
We also plot the function $\mathcal{A}(r, u, \theta)$ of \eqref{Adefsp}
in  Fig.\ref{A} (b).

Besides, we also need an appropriate boundary condition on some domain.
Ref.\cite{marin2011}  has obtained  the  redshift-space 3PCF
of luminous red galaxies of ``DR7-Dim"
(61,899 galaxies in the  range $0.16 \leq z \leq 0.36$)
from  SDSS.
In Figure 6 and Figure 7 of Ref.\cite{marin2011},
the reduced  $Q(s, u, \theta)$ are given in the domain
\[
 s=[7.0 , 30.0]\, h^{-1} {\rm Mpc}, ~~  \theta =[0.1 , 3.04]
\]
at five respective values  $s=7,10,15,20,30\, h^{-1}$Mpc at a fixed $u=2$.
Specifically,
we shall use the measured $Q(s, u, \theta)$ at $s=7 h^{-1}$Mpc  and $s = 30\, h^{-1}$Mpc
as a part of the boundary condition, which is fitted by
\ba \label{border2011b}
Q(\theta)=
\begin{cases}
	1.6563 + 56.8042 \, \theta -
	16.7962 \, \theta^2 + 6.7985 \, \cos \theta \\
	\hspace*{0.4cm}	- 6.8108 \, \cos 2 \theta
	- 0.4031 \, \cos 3 \theta
	- 54.9452 \, \sin \theta \\
	\hspace*{0.4cm}- 2.088 \, \sin 2 \theta
	+ 0.7494 \, \sin 3 \theta,
        ~~~~~~(s=7.0 h^{-1} {\rm Mpc}) \\
	86.5647 + 1040.2889 \, \theta
	- 320.5828 \, \theta^2 + 53.4609 \, \cos \theta \\
	\hspace*{0.4cm}	- 136.5958 \, \cos 2 \theta
	-2.3371 \, \cos 3 \theta
	- 1049.9285 \, \sin \theta  \\
	\hspace*{0.4cm}	- 14.6843 \, \sin 2 \theta
	+ 17.0408 \, \sin 3 \theta,
        ~~~~~~ (s=30.0 h^{-1} {\rm Mpc})\\
\end{cases}
\ea
Also  from Figure 6 and Figure 7  of Ref. \cite{marin2011},
we give the fitted $Q(s, u, \theta)$ at $\theta=0.1$ and $\theta=3.04$
as  another part of the boundary  condition
\ba \label{border2011a}
Q(s)=
\begin{cases}
	0.8979 + 0.03968 \, s - 0.00035 \, s^2 , &(\theta=0.1)	\\
	1.607 - 0.08998 \, s + 0.004731 \, s^2, &(\theta=3.04) . \\
\end{cases}
\ea
\eqref{border2011b} and \eqref{border2011a} lead to
the  boundary values of $\zeta(s, u, \theta)$ on the domain,
by virtue of  the relation   (\ref{grothpeeblesansatz}).
The redshift distance $s$ is used in Ref.\cite{marin2011}
which may differ from the real distance $r$ due to the peculiar velocities.
We shall neglect this error in our computation.
To match  the observational data \cite{marin2011},
the parameters are chosen as the following:
$a_r=-1043.8$, $b = -1627.3$,
$c = -36.4$,
$g=-5586.6$,
$R_a=1.66$,  $R_b = -0.34$, $Q= 1.1$,
$k_J=0.161 \, h \mathrm{Mpc}^{-1}$.

Eq.(\ref{3PCF_02sp}) is a convection-diffusion partial differential equation,
and we employ the streamline diffusion method \cite{Elman}
to solve it numerically.
We obtain the solution $\zeta(r, u, \theta)$
and   the reduced $Q(r, u, \theta)$ by the relation   (\ref{grothpeeblesansatz}).

\begin{figure}[htb]
\centering
        \includegraphics[width = .95\linewidth]{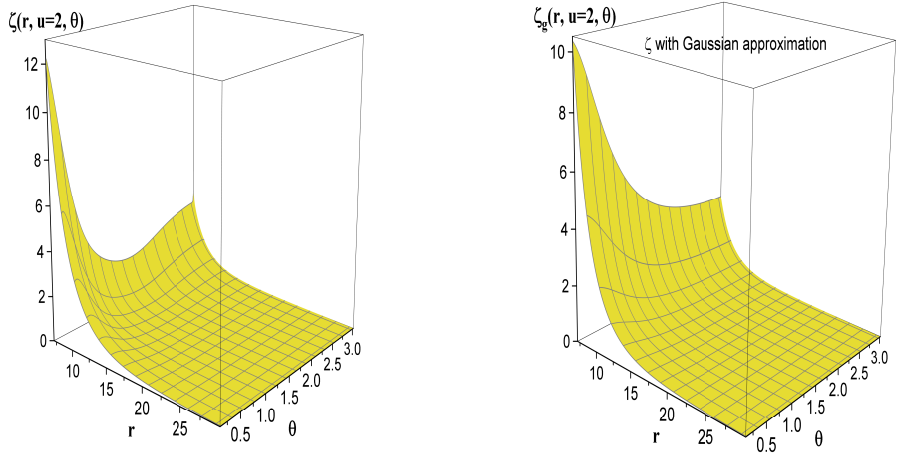}
\caption{ (a): The solution $\zeta(r, u, \theta)$
          shows a shallower  $U$-shape along  $\theta$,
          and decreases  monotonously along  $r$.
         (b): The Gaussian solution $\zeta_g(r, u, \theta)$ of eq.\eqref{gauss3pt}
            decreases  monotonously along both $\theta$ and $r$.
}
\label{newzeta3d}
\end{figure}

Fig.\ref{newzeta3d}  (a)  plots the  surface of  $\zeta(r, u, \theta)$
as a function of ($r,\theta$),
which exhibits a shallow $U$-shape along $\theta$
and turns up  at $\theta \gtrsim \pi/2$.
This feature of solution is consistent with observations \cite{Guo2013,Guo2016}.
$\zeta(r, u, \theta)$ decreases monotonously along $r$
up to  $30 h^{-1}$Mpc.
The highest  values of $\zeta(r, u, \theta)$ occur at small $r$ and $\theta$.
For a comparison,
Fig.\ref{newzeta3d} (b) plots the Gaussian solution  $\zeta_{g}(r, u, \theta)$
of eq.\eqref{gauss3pt},
which decreases monotonously along both $\theta$ and $r$,
having no $U$-shape along $\theta$.

\begin{figure}[htbp]
	\centering
	\includegraphics[width=0.8\columnwidth]{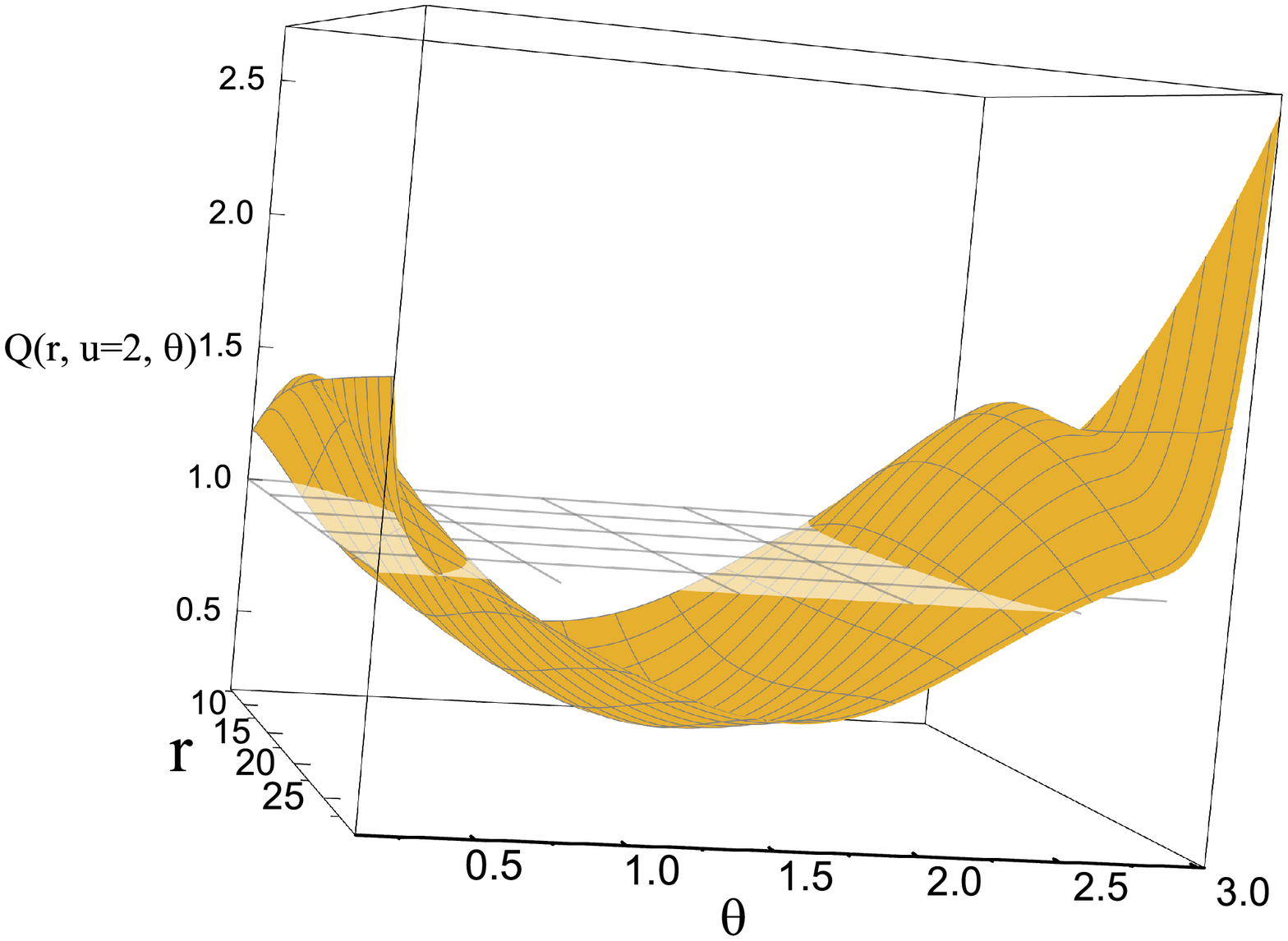}\\
	\caption{  \label{q3d}
The surface of $Q(r, u, \theta)$
deviates from the Gaussianity plane $Q(r, u, \theta)=1$,
 exhibits a deeper $U$-shape along $\theta$, and varies weakly along the radial $r$.
	}
\end{figure}

Fig.\ref{q3d} plots the surface of reduced $Q(r, u, \theta)$ as a function of ($r,\theta$),
which deviates  from the Gaussianity plane $Q(r, u, \theta)=1$,
exhibits a deeper $U$-shape along $\theta$, and varies along the radial $r$.
The highest values of $Q(r, u, \theta)$ occur at large $r$ and $\theta$,
just opposite to $\zeta(r, u, \theta)$.
The variation along $r$ is comparatively weaker than the variation along $\theta$.
These features  are consistent
with observations \cite{marin2011,McBride2011a,McBride2011b}.

\begin{figure}[htbp]
	\centering
	\includegraphics[width=0.8\columnwidth]{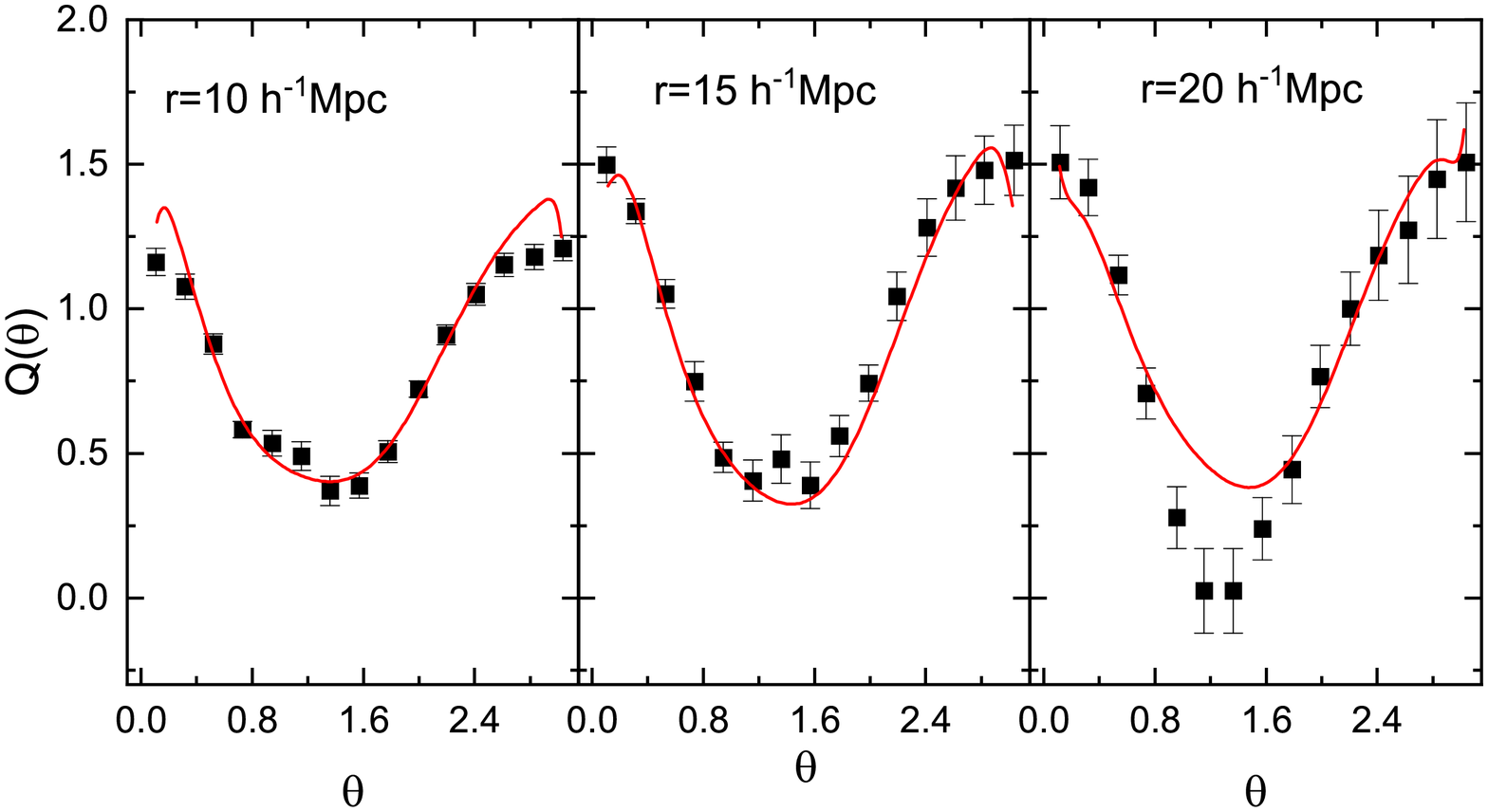}\\
	\caption{  \label{qcompare}
The solid line:  $Q(r, u, \theta)$ at $u=2$
 converted from the  solution $\zeta(r, u, \theta)$.
The points:  the SDSS observational data from Fig.6 and Fig.7 of Ref. \cite{marin2011}.
Three plots are for $r=10  h^{-1} {\rm Mpc}$,
$15  h^{-1} {\rm Mpc}$, $20  h^{-1} {\rm Mpc}$, respectively.
$Q(r, u, \theta)$ deviates from $Q(r, u, \theta)=1$ of Gausianity
and forms a $U$-shape along the elevation angle $\theta=[0,3]$,
  agreeing  with the  data.
	}
\end{figure}

To compare with observations,
Fig.\ref{qcompare}  shows    $Q(r, u, \theta)$ as a function of $\theta$
at  respectively fixed  $r=10, 15 ,20  \,  h^{-1} {\rm Mpc}$.
$Q(r, u, \theta)$ agrees well with the data  of Ref.\cite{marin2011}
available  in the range $\theta=(0.1 \sim 3.0)$.

\begin{figure}[htbp]
\centering
\includegraphics[width=0.8\columnwidth]{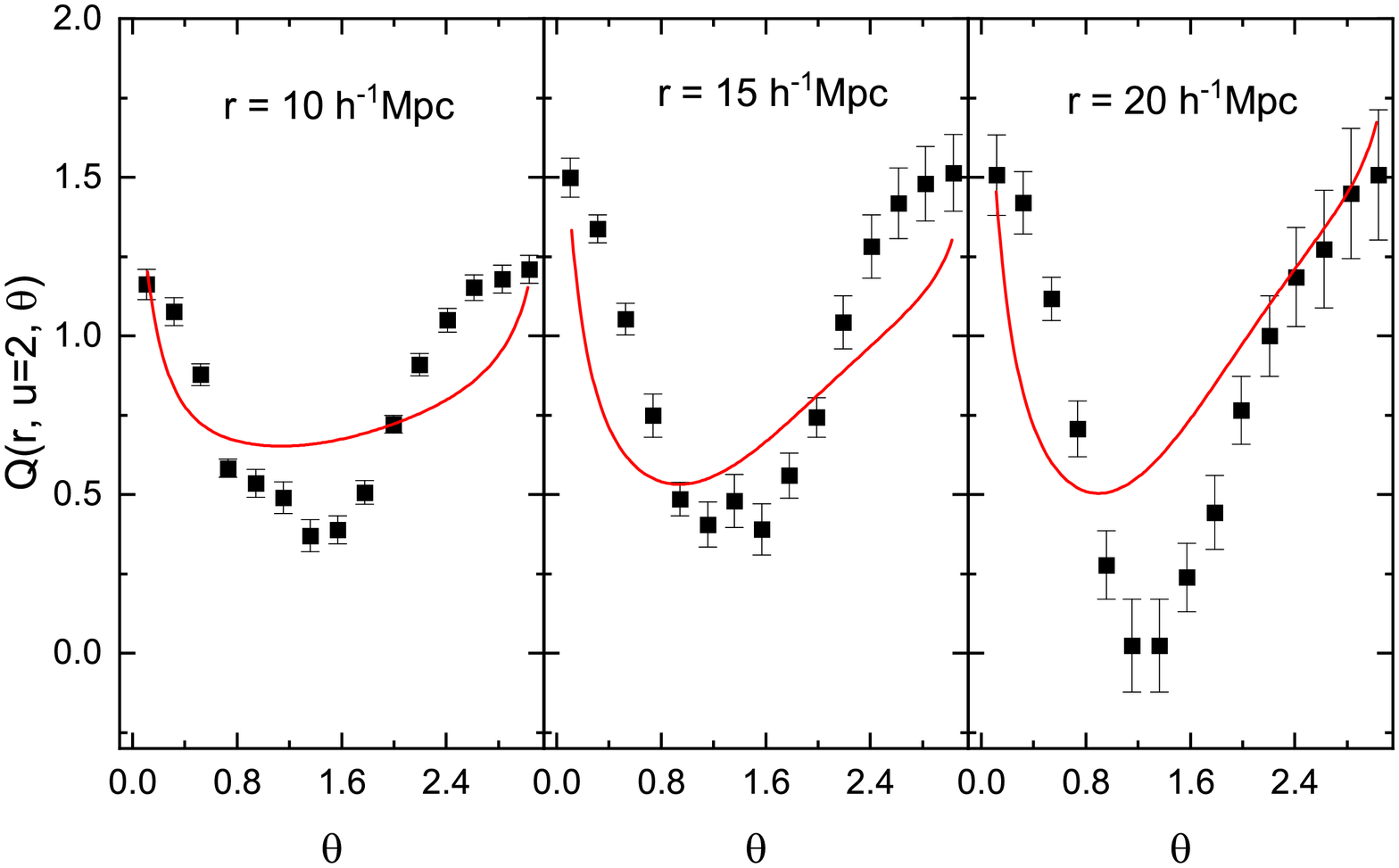}
\caption{ Similar to Fig.\ref{qcompare}.
 $Q(r, u, \theta)$ is plotted, using another set of parameters:
$k_J=0.12822 \,h^{-1}$Mpc, $a_r= 34.03 \, k_J$, $b=3.36$, $c=1.8844$,  $g = 1  + c/(4 k_J^2)$,
$R_a=-2.06$, $R_b=6.64$, $Q=0.7$.
The fitting to the data is not as good as  Fig.\ref{qcompare}.
}
\label{qNew}
\end{figure}

As an example,
Fig.\ref{qNew}  plots  $Q(r, u, \theta)$ with another set of parameter values,
and the fitting is not as good as that in Fig.\ref{qcompare}.

\section{Conclusions and Discussions}

We have presented an analytical study the 3-point correlation function of galaxies
based on the field theory of density fluctuations of a Newtonian gravitating system,
and have derived the nonlinear field equation \eqref{3PCF}
of  $G^{(3)}$
up to the second order  density fluctuation.
This work is a continuation of  the previous works
on the 2PCF \cite{Zhang2007,zhang2009nonlinear,ZhangChen2015}
and on the Gaussian 3PCF \cite{ZhangChenWu2019} .

By adopting the Fry-Peebles  ansatz
to deal with the 4PCF,
and the Groth-Peebles ansatz to deal with the squeezed  3PCF, respectively,
we have made  eq.\eqref{3PCF} into
the closed equation \eqref{3PCF_02} of $G^{(3)}$,
equivalently eq.\eqref{3PCF_02sp} of $\zeta$ in spherical coordinate.
For  coherency,
we have used the observed 2PCF and the boundary condition from SDSS DR7 \cite{marin2011},
in solving for the  3PCF.

The solution $\zeta(r, u, \theta)$ exhibits a shallow $U$-shape along $\theta$,
agreeing with the observed one.
And, nevertheless, $\zeta(r, u, \theta)$ decreases monotonously along $r$,
at least up to $30\, h^{-}$Mpc of the domain in our computation.
For comparison, we also plot the Gaussian solution $\zeta_{g}(r, u, \theta)$,
which decreases monotonously along both $\theta$ and $r$, having no $U$-shape along $\theta$.
The difference between $\zeta$ and $\zeta_g$
implies the non-Gaussianity of the distribution of galaxies.

The  non-Gaussianity is directly indicated  by the reduced $Q(r, u, \theta)$.
The solution  $Q(r, u, \theta)$ deviates from the Gaussianity plane $Q(r, u, \theta)=1$,
also exhibits a $U$-shape along $\theta$, just like $\zeta(r, u, \theta)$,
agreeing  with the observations \cite{marin2011}.
In fact, by its definition \eqref{grothpeeblesansatz},
$Q(r, u, \theta)$ shares the same $\theta$-dependence as $\zeta(r, u, \theta)$,
and its denominator consists of $\theta$-independent  $\xi(r)$.
Along $r$, however,  $Q(r, u, \theta)$ varies non-monotonically, scattering around $1$,
unlike  $\zeta(r, u, \theta)$.
Moreover, the highest values of $Q(r, u, \theta)$ occur at large $r$ and $\theta$,
a behavior just opposite to $\zeta(r, u, \theta)$.
These two features of $Q(r, u, \theta)$
are due to the behavior of $\xi(r)$ which is large at small $r$
and suppresses $Q(r, u, \theta)$ thereby.

This preliminary study of 3PCF in this paper should be extended,
and several issues need more investigation in future,
such as the impact of physical parameters,
exploration of parameter space  in association with 2PCF,
and the  effect of cosmic expansion.

\section*{Acknowledgements}

Y. Zhang is supported by
NSFC Grant No. 11675165,  11633001,   11961131007.

\end{document}